\documentclass[journal]{IEEEtran}
\usepackage{cite}
\usepackage{graphicx}
\usepackage{booktabs}
\usepackage{tikz}
\usepackage{eurosym}
\usepackage{alphalph,etoolbox}
\usepackage{url}
\usepackage{epstopdf}
\usepackage{amsfonts,amsmath,amsthm,amssymb}
\usepackage{subfigure}
\usepackage{stfloats}
\usepackage{eurosym}
\usepackage{enumerate}
\usepackage{color}
\usepackage{arydshln}
\usepackage{longtable}
\usepackage{mathrsfs}
\usepackage{bm}
\usepackage[T1]{fontenc}
\usepackage[latin1]{inputenc}
\usepackage[shortlabels]{enumitem}
\usepackage{tikz}
\usepackage{footnote}
\makesavenoteenv{minipage}
\makesavenoteenv{enumerate}
\usepackage{endnotes}
\usetikzlibrary{arrows}

\usepackage{csquotes}

\usepackage{psfrag}

\patchcmd{\subequations}{\alph{equation}}{\alphalph{\value{equation}}}{}{}

\include{def}

\begin{document}

\graphicspath{{Figures/}}

\title{A hybrid green energy-based framework with a multi-objective optimization approach for optimal frost prevention in horticulture}


\author{Ercan Atam, Tamer F. Abdelmaguid,  Muhammed Emre Keskin, Eric C. Kerrigan
\thanks{
Ercan Atam and Eric C. Kerrigan are with Electrical and Electronic Engineering Department of Imperial College London, United Kingdom. Kerrigan is also with the Department of Aeronautics.

Tamer F. Abdelmaguid is with
Department of Mechanical Design \& Production, Cairo University, Egypt.

Muhammed Emre Keskin is with
Department of Industrial Engineering,
Ataturk University, Turkey.

E-mails: e.atam@imperial.ac,uk,
tabdelmaguid@eng.cu.edu.eg,
emre.keskin@atauni.edu.tr,
e.kerrigan@imperial.ac.uk.
}}

\markboth{IEEE Transactions on Emerging Topics in Computing}%
{
}

\maketitle

\begin{abstract}
In this paper, first we propose a novel hybrid renewable energy-based
solution for frost prevention in horticulture applications
involving active heaters.
Then, we develop a multi-objective robust optimization-based formulation
to optimize the distribution of a given number of active heaters in a given large-scale orchard. The objectives are to optimally heat the orchard
by the proposed frost prevention system and to minimize the total length of the energy distribution pipe network (which is directly related to the installation cost and the cost of energy losses during energy transfer). Next, the resulting optimization problem is approximated using a discretization scheme.
A case study is provided to give an idea of the potential savings using the
proposed optimization method compared to the result from a heuristic-based  design, which showed a 24.13\% reduction
in the total pipe length and  a 54.29\% increase in
frost prevention.

\end{abstract}

\begin{IEEEkeywords}
Hybrid energy systems, active horticulture frost prevention systems,
optimal design, multi-objective-optimization, k-MST, robust optimization.
\end{IEEEkeywords}
\IEEEpeerreviewmaketitle

\section{Introduction}

Fruit constitutes an indispensable component of our daily nourishment
and a significant fraction of export income for many countries (for example, $\approx$ \euro 15-20 billion
for EU member states \cite{EU_fruit_vegetable_sector_2020}). However,
during budding and/or flowering periods, buds or flowers of many fruits are vulnerable to low temperature levels, especially to temperatures below $0 ^{\circ}$C,
and hence for them frost occurrence can cause a significant yield loss.
As an example, the frost occurrence on March 31, 2014 in Malatya, Turkey
(which is the world capital for dry apricot production) caused
around 95\%  of yield loss whose economic value was \$1.2 billion US \cite{Malatya_IGTHM_2016}.

Active heaters which blow hot air on trees (to be called  hot air blowing active
heaters (HABAHs) in the rest of the paper) during frost periods
can be a  good choice for frost prevention in large-scale
orchards \cite{Atam_Arteconi_2017,Atam_et_al_2020}.
However, a significant barrier for widespread use of such systems
is the installation and operation costs. Adoption of renewable
or hybrid energy  solutions (combination of renewable and non-renewable energies) to create hot
air can reduce these costs up to some level. However, independent of the used energy source,
two important design problems for active heater-based frost prevention systems are  (i) the optimal distribution of a given number of
active heaters inside a large-scale orchard so that maximum protection
against frost episodes will be achieved and (ii) if hot air is distributed to HABAHs through a piping network, then optimization of the distribution pipe network to reduce pipeline cost and thermal energy losses.

\begin{figure*}[h!]
\centering
\includegraphics[scale=0.29]{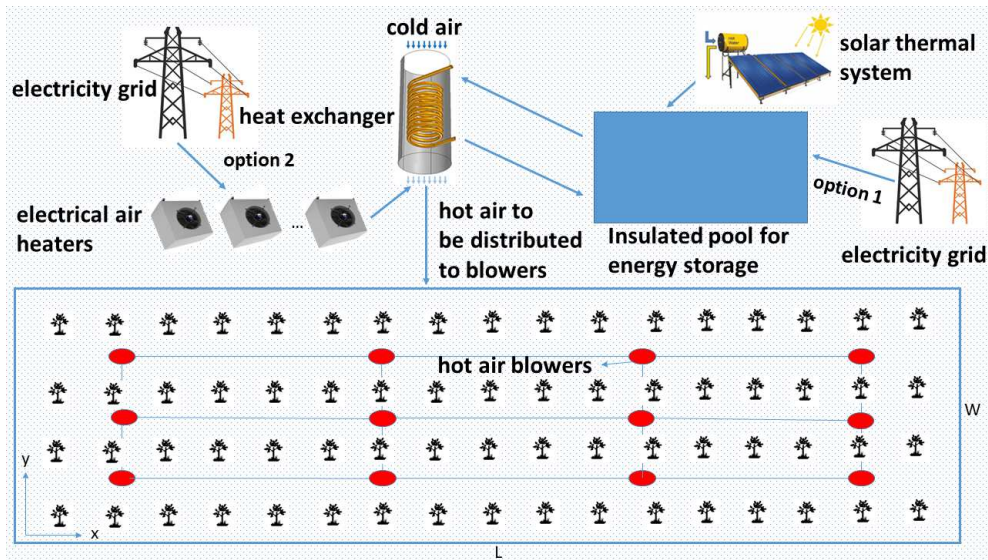}
\caption{Schematic of a hybrid energy-based frost prevention system integrating HABAHs for large-scale fruit orchards.}
\label{fig:rost_prev_energy_system_new}
\end{figure*}

A simplified schematic of the proposed hybrid energy-based
frost prevention system where HABAHs are used is given in Figure \ref{fig:rost_prev_energy_system_new}.
This system is called a hybrid energy system because a combination
of renewable and non-renewable energy sources is used.
The working principle of the system is as follows.
Solar energy through solar collectors  will be used to heat water and store it inside an insulated pool. The stored hot water
will be used to heat air via a water-air heat exchanger
and the hot air will be distributed through a pipe network
to a number of blowers inside the orchard which
will blow hot air on trees during frost periods. When necessary, additional
energy from the grid will also be used to heat water in the pool (option 1),
or to directly heat air via an array of electrical air heaters (option 2).

The use of a solar thermal system can provide significant advantages for regions with rich solar radiation,
and for such regions, the extra energy that may be needed from the electricity grid to further heat the water or air will be a small fraction of the total energy used for frost prevention.
As a result, the proposed hybrid solution will be an economically feasible and cost-effective solution. The economic feasibility analysis of the proposed hybrid frost prevention system (such as calculation of investment/operational costs and payback time)
is not in the scope of the current paper and this will be studied in a future paper.

Note that the use of solar thermal collectors and the different insulation designs have already
been considered for heating  swimming  pools
with promising results \cite{Francey_et_al_1980, Dongellini_et_al_2015}.
In the proposed hybrid energy solution, the  purpose of the pool is to  store solar thermal energy
for a different application and pool will be completely covered from all sides with the most effective insulation material
to minimize energy losses. As a result, the insulation and hence the
storage efficiency requirements of the solar thermal-driven pool system
of this study are  more demanding.

Next, we developed a  multi-objective robust optimization-based approach for  optimal placement
of a number of HABAHs inside a given large-scale orchard
for effective frost prevention and optimal design of the energy distribution network of the proposed framework. The basic building blocks for the developed optimization scheme
consists of (i) assuming a physically-reasonable function for spatial variation of the  heating effect of a heater,
(ii) modeling the optimal heating of the orchard  against frost
as an optimization constraint,
(iii) constructing a k-node minimum spanning tree (k-MST)
from a large undirected graph with unknown edge weights
for optimal design of the layout of the energy distribution pipe network to reduce installation and energy loss costs. The resulting problem is a large-scale
multi-objective robust mixed-integer nonlinear programming
problem where we use a discretization scheme
to approximate the  problem  with a mixed-integer linear programming (MILP) problem.
Moreover, we developed a MILP-based k-MST formulation which is very useful for multi-objective optimization
problems for which k-MST is a part. The k-MST is known to be NP-complete~\cite{Fischetti1994}. Furthermore,
suboptimal k-MST heuristics developed in the literature,
such as \cite{Arya_Ramesh_1998, Arora_Karakostas_2006, Garg_1996, Garg_2005},
are not found suitable for the considered application since
multiple objective functions are studied simultaneously, in addition to other non-traditional constraints that should be satisfied.

This paper is the
first attempt in the open literature to propose
a novel hybrid energy-based solution to the frost prevention problem
in large-scale fruit orchards. This is also a pioneering work in
proposing a multi-objective robust optimization-based approach,
including a novel MILP-based k-MST formulation to tackle the challenging, but important inherent optimal design problems
in applications based on a k-MST.

The rest of the paper is organized as follows.
In Section~\ref{sec:Spatial heater power effect variation modeling},
an empirical, but  physically reasonable model for the spatial
variation of the heating effect of a HABAH is given.
The development of the multi-objective robust optimization  formulation
for the considered application is described in  detail in Section \ref{sec:Energy distribution network optimization formulation},
which includes  the proposed discretization scheme for optimization approximation, the k-MST and robust counterpart formulations, each of which is a part of the overall optimization problem.
A case study is given in Section \ref{sec:A case study} to demonstrate the
optimal savings using the optimization-based design compared to a heuristic-based design.
Finally, the main findings of this study along with some
future research directions are given in Section \ref{sec:Conclusions}.

\section{Spatial heater power effect variation modeling}
\label{sec:Spatial heater power effect variation modeling}

The heating effect of any hot air blower-type active heater decreases
with distance and  this effect depends on a number of factors
such as the installation configuration of the air blower,
the mass flow rate of blown air and its temperature.
In this paper, we assume the following representative empirical
function for the spatial variation of the heating effect of a given  HABAH:

\begin{align}
P_{x_i, y_i}(x, y)=& P_0 k_i^u\underbrace{e^{-k_{tun}\sqrt{(x-x_i)^2+(y-y_i)^2}}}_{\triangleq f(x,y;x_i,y_i)} \nonumber \\
               =& P_0 k_i^u f(x,y;x_i, y_i)
\end{align}
\noindent where $f(x,y;x_i,y_i)$ is the ``effective heating power" of the $i$-th active heater, which is centered at $(x_i, y_i)$, at the point $(x, y)$. Basically, $f(x,y;x_i, y_i)$ reflects the fraction of the maximum  heating power $P_0$ transferred to the point $(x,y)$ in the orchard.
The parameter $k_{tun}$ is a tuning parameter  to vary the heating effect
and the parameter $k_i^u$ is an uncertain parameter,
lying in the interval $[\underline{k_i^u}, \, \overline{k_i^u}]$,
to account for uncertainty in $P_{x_i, y_i}(x, y)$.
For a specific type of hot air blowing heater, a function similar to  $P_{x_i, y_i}(x, y)$ can be used, and hence the developed  general optimization framework
in the next section can be used for any hot air blower.

\section{Energy distribution network optimization formulation}
\label{sec:Energy distribution network optimization formulation}

\subsection{The optimization problem}

The motivation behind the use of optimization for the presented application is as follows. (i) For a given orchard, we want to locate $k$ HABAHs
inside the orchard in order to
heat the given orchard optimally in a balanced way. (ii)
The heaters are connected through a pipe network in such a way that
the length of the energy distribution pipeline network
is minimized. Minimization of this length has a
twofold benefit: first, the installation cost is minimized
and, second the energy losses from the pipe network
during energy circulation is minimized (the shorter the total pipe length, the smaller is the thermal energy loss because heat loss increases linearly with  total pipe length).

As constraints of the optimization problem, the following conditions should be satisfied:

\begin{align}
&  \underline{f}-\mu_{s}^l \le \displaystyle \sum_{i=1}^{k}k_i^uf(x_s^{cp},y_s^{cp}; x_i,y_i) \le \overline{f}+\mu_{s}^u,\, s=1,\cdots,n_{cp} \label{power_fr_const} \\
& (x_i-x_j^t)^2+ (y_i-y_j^t)^2 \ge d_{ht}^2, \quad  i=1,\cdots,k,\, j=1,\cdots,n_t
\label{distance_from_heater_to_tree_constraint}
\end{align}
Here, \eqref{power_fr_const} is used to enforce the condition
that at each selected discrete check point
$(x_s^{cp},y_s^{cp})$
in the orchard (``cp" meaning check point) the sum
of power fractions from all heaters should be in the range $[\underline{f}, \overline{f}]$
whenever possible (if not possible, then minimum violations $\mu_{s}^l, \mu_{s}^u$
are allowed);
\eqref{distance_from_heater_to_tree_constraint} is used to ensure that the distance between a heater and
the root of a tree is a minimum of $d_{ht}$ meters because heaters should not be installed in
areas occupied by tree stems and branches (assuming that the area occupied by a tree is approximately
a circular area with center $(x_j^t, y_j^t)$ and radius $d_{ht}$).
Note that the constraints  in \eqref{power_fr_const} are uncertain
constraints since  $k_i^u$s are uncertain parameters,
lying in the interval $[\underline{k_i^u}, \, \overline{k_i^u}]$.

The cost function to be minimized is
\begin{align} \label{cost_function}
\sum_{(i,j) \in V \times V,\, j > i}q_{ij}\sqrt{(x_i-x_j)^2+(y_i-y_j)^2}+\alpha\sum_{s=1}^{n_{cp}}(\mu_s^l+\mu_s^u)
\end{align}
where $V=\{1, 2, \cdots, k\}$,
$q_{ij}$ is a binary variable indicating whether the energy pipe network contains a ``direct" pipe branch ($q_{ij}=1$) or not ($q_{ij}=0$) from heater $i$ to heater~$j$.
The objective function consists of the sum of two terms
where the first one (the summation term) denotes the length of the minimum spanning tree consisting of k nodes (k-MST) and
the second term is used to penalize power range violations at check points.

Note that the above optimization problem consisting of
the cost function \eqref{cost_function}, constraints
\eqref{power_fr_const}-\eqref{distance_from_heater_to_tree_constraint}
and k-MST constraints (which we did not write at this point since
they will be developed later in detail) is a multi-objective
robust nonlinear programming problem.

\subsection{Discretization of orchard domain}
\label{subsection:Approach 2: discretization of nonlinear terms}

In this section, we propose a discretization-based
approach to be used in solving approximately the multi-objective
robust nonlinear programming problem given in the previous section.
In this approach we create a set of uniform discrete points inside the orchard
satisfying the constraints in \eqref{distance_from_heater_to_tree_constraint}
as candidate heater location points (ch-lps) to place the heaters and we denote this set by
$\mathcal{V}$ with $|\mathcal{V}|=n_{ch-lps} \gg k$. Next, we construct an undirected weighted graph $\mathcal{G}=(\mathcal{V},\mathcal{E})$
where $\mathcal{E}$ is the set of weighted edges
between each candidate heater location point and the edge weights $d_{e}^{ch-lps}$,
$e \in \mathcal{E}$, are the distance between the candidate heater location points.
The advantage of  using this discretization
technique is that the nonlinear constraints in
\eqref{distance_from_heater_to_tree_constraint}
will be eliminated from the optimization problem, and we will be able to replace the nonlinear terms in \eqref{power_fr_const} and \eqref{cost_function}
with linear terms as shown later.

Since k-MST is a part of the considered multi-objective robust optimization
problem, next we develop a mixed-integer linear programming
formulation for the k-MST problem.

\subsection{MILP-based formulation of k-MST problem}
In this section, we extend the original Miller-Tucker-Zemlin (MTZ) MILP model developed for the travelling salesman problem~\cite{MTZ1960} to the k-MST problem.

\subsubsection{Model structure and main variables}
Consider a generic undirected graph $\mathcal{G}=(\mathcal{V},\mathcal{E})$ where
$\mathcal{V}$ is the node set and $\mathcal{E}$ is the set of weighted undirected edges. For every edge $e = \{i,j\} \in \mathcal{E}$, where $i, j \in \mathcal{V}$, a binary decision variable $z_e$ is defined which represents the edge's inclusion (value of 1)/exclusion (value of 0) in the k-MST. Furthermore, for every node $i$, we define a binary variable $\ell_i$ that equals 1 if node~$i$ is included in the k-MST and zero otherwise. It is required to construct a tree with exactly $k$ nodes to achieve the stated objectives. Next, we develop the necessary constraints of the model.

\subsubsection{Constraints}
The MTZ formulation is based on defining a pair of binary variables
for each edge that suit the directed traveling salesman tour, denoted as $w_{(i,j)}$ and $w_{(j,i)}$. The following set of constraints represent their relationships with $z_e=z_{\{i,j\}}$:
\begin{equation}
z_e = w_{(i,j)} + w_{(j,i)} \;\;\;\; \forall e = \{i,j\} \in \mathcal{E} \label{KMST_1}
\end{equation}

The relationships provided in~\eqref{KMST_1} enable smooth translations between the decisions of including/excluding the hypothetical directed arcs $(i,j)$ and $(j,i)$ and the inclusion/exclusion of the undirected edge $\{i,j\}$.

In similar minimum spanning tree formulations, a node in $\mathcal{V}$ is arbitrarily selected and labeled as the terminal node. All resultant directed paths using the hypothetical directed arcs should end at that terminal node as part of the restrictions that lead to the formation of a tree~\cite{Abdelmaguid2018}. In the case of k-MST, since not all the nodes will be included in the tree, that terminal node cannot be chosen from the nodes in $\mathcal{V}$. Therefore, we add a dummy node $\tau$ to represent the terminal node in the current formulation. We also define a set of dummy edges $D = \{\{i,\tau\}: i \in \mathcal{V}\}$. The lengths of the edges in $D$ do not affect the objective functions, and therefore, their values are not of concern. For every dummy edge $e \in D$, a binary decision variable $z_e$ is augmented to the model, as well as pairs of $w_{(i,j)}$ and $w_{(j,i)}$ binary variables associated with its corresponding hypothetical directed arcs. Accordingly, similar to constraints~\eqref{KMST_1}, the following constraints are added to the model.
\begin{equation}
z_e = w_{(i,\tau)} + w_{(\tau,i)} \;\;\;\; \forall e = \{i,\tau\} \in D
\label{KMST_4}
\end{equation}

In the MTZ formulation, there should be exactly one arc directed out of node $i$, as well as exactly one arc directed into it in order to complete the travelling salesman tour. This restriction is not suitable for trees, since a node in a tree can have more than two edges connecting it to more than two nodes. As demonstrated in~\cite{Abdelmaguid2018}, this can be circumvented in the minimum spanning tree formulation by allowing only one restriction. That is, having exactly one arc directed out of a node, except the terminal node $\tau$. In the k-MST formulation, this has to be governed by the condition of whether this node is included in the tree or not. The following constraints, represent these conditions:
\begin{align}
& \sum\limits_{j \in \mathcal{V} \cup \{\tau\}, j \neq i}{w_{(i, j)}} = \ell_i & \forall i \in \mathcal{V} \label{KMST_2} \\
& \sum\limits_{j \in \mathcal{V}, j \neq i}{w_{(j, i)}} \leq (k-1) \ell_i & \forall i \in \mathcal{V}
\label{KMST_3}
\end{align}

Here, the constraints in \eqref{KMST_2} restrict the number of selected outgoing arcs starting at node $i$ to be exactly 1 if it is included in the k-MST, and to be zero otherwise, for all nodes $i \in \mathcal{V}$. Meanwhile, the constraints in \eqref{KMST_3} make sure that node $i$ will be connected by incoming arc(s) only when it is selected to be included in the k-MST.

Constraints~\eqref{KMST_2} and~\eqref{KMST_3} will result in a set of paths that start at a subset of nodes and can intersect at intermediate nodes. In the current model, all such paths should end at the dummy terminal node ($\tau$). To achieve that, the following two constraints are added:
\begin{equation}
\sum\limits_{j \in \mathcal{V}}{w_{(\tau, j)}} = 0
\label{KMST_5}
\end{equation}
\begin{equation}
\sum\limits_{i \in \mathcal{V}}{w_{(i, \tau)}} = 1
\label{KMST_6}
\end{equation}

Constraints~\eqref{KMST_5} and~\eqref{KMST_6} make sure that only one edge connecting node $\tau$ will appear in the final solution. This restriction is necessary to make sure that all resultant $k$ nodes will be connected. The only edge that connects node $\tau$ to one of the other $k$ nodes can then be excluded when the final MILP solution is interpreted.

Subtours in the MTZ formulation are eliminated by introducing continuous variables $u_i$ for each node $i \in V \cup \{\tau\}$. The elimination is done by allowing a directed arc $(i,j)$ to appear in the solution only when $u_i > u_j$. The following constraints maintain this logic:
\begin{equation}
u_i \geq u_j + w_{(i, j)} - k (1 - w_{(i, j)}) \;\;\; \forall i \in \mathcal{V}-\{j\} \;\; \forall j \in \mathcal{V} \cup \{\tau\}
\label{KMST_7}
\end{equation}

The range of values that can be assigned to the $u_i$ variables are defined by the following constraints:
\begin{align}
& u_{\tau} = 0 & \label{KMST_8} \\
& u_i \leq (k - 1) \ell_i & \forall i \in \mathcal{V} \label{KMST_9} \\
& u_i \geq \ell_i & \forall i \in \mathcal{V}  \label{KMST_10}
\end{align}

Finally, the constraint that specifies the number of selected nodes to be exactly $k$ and the domain constraints are defined as
\begin{align}
& \sum\limits_{i \in V}{\ell_i} = k & \label{MST_11} \\
& \ell_i \in \{0, 1\} & \forall i \in \mathcal{V} \label{MST_12} \\
& w_{(i, j)}, z_{\{i,j\}} \in \{0, 1\} & \forall i,j \in \mathcal{V} \cup \{\tau\} \label{KMST_13}
\end{align}

\subsection{Robust counterpart formulation for uncertain constraints}
\label{subsec:Robust counterpart formulation}

As mentioned before, the constraints in  \eqref{power_fr_const} are uncertain
constraints since  the $k_i^u$ are uncertain parameters
lying in the interval $[\underline{k_i^u}, \, \overline{k_i^u}]$.
To develop a deterministic MILP problem corresponding to the robust MILP, which is called  the ``robust counterpart" in the robust optimization
literature \cite{Bertsimas_Sim_2004, Li_et_al_2011},
we need to express the uncertain constraints \eqref{power_fr_const}
in a deterministic form such that they are satisfied for all realizations of the uncertain parameters $k_i^u,\, i=1,\cdots, k$ in their range $ \underline{k_i^u}\le k_i^u \le \overline{k_i^u}$. This happens
if in \eqref{power_fr_const} we replace $ k_i^u$s with their lower bounds $\underline{k_i^u}$ and upper bounds $\overline{k_i^u}$
for "$\le$" and "$\ge$", respectively.
Using this replacement combined with
the discretization scheme, we obtain the ``robust counterpart constraints"
as
\begin{subequations}
\begin{align}
&  \underline{f}-\mu_{s}^l \le \displaystyle \sum_{i=1}^{n_{ch-lps}}\ell_i\underline{k_i^u}h_{is}  & s=1,\cdots,n_{cp}\\
& \displaystyle \sum_{i=1}^{n_{ch-lps}}\ell_i\overline{k_i^u}h_{is} \le \bar{f}+\mu_{s}^u & s=1,\cdots,n_{cp}
\end{align}
\end{subequations}
where $h_{is} \triangleq f(x_s^{cp},y_s^{cp}; x_i^{ch-lp},y_i^{ch-lp})$ denotes
heat influence of a heater located at the $i$-th candidate heater location point with coordinates  $(x_i^{ch-lp},y_i^{ch-lp})$ at the
$s$-th check point with  coordinates  $(x_s^{cp},y_s^{cp})$.

\subsection{The overall approximated optimization problem}

Collecting all the previous developments, we can express
the overall multi-objective robust optimization problem as in
\eqref{approx_MILP}. The
parameters $\beta_1^{nor},\beta_2^{nor}$
are used to normalized each part of the
cost function to lie in the interval [0,1].
For a given large-scale orchard, it is important that a proper number of candidate heater location points
and check points are created so that a good trade off is achieved between
the optimality of the resulting approximated MILP problem and the number of
resulting associated binary variables, which affect the solvability and
the time required for the solution of the approximate MILP.

\begin{figure*}[h!]
\hrule
\vspace{0.1cm}
Approximate MILP:
\vspace{0.1cm}
\hrule
\begin{subequations} \label{approx_MILP}
\begin{eqnarray}
&\hspace{-1cm} \displaystyle \min_{z_{\{i,j\}}, w_{(i,j)}, w_{(j,i)},  w_{(i,\tau)}, w_{(\tau,i)}, \ell_i, u_i, \mu_s^l, \mu_s^u}\left\{\frac{1}{\beta_1^{nor}}\sum_{\{i,j\} \in \mathcal{E} }z_{\{i,j\}}d_{i,j}^{ch-lp}+\frac{\alpha}{\beta_2^{nor}}\sum_{s=1}^{n_{cp}}(\mu_s^l+\mu_s^u) \right\} &  \label{approx_MILP_obj_func}\\
& \eqref{KMST_1}-\eqref{KMST_13} & (\text{k-MST constraints}) \label{approx_MILP_kMST_cons} \\
&  \underline{f}-\mu_{s}^l \le \displaystyle \sum_{i=1}^{n_{ch-lps}}\ell_i\underline{k_i^u}h_{is}  & s=1,\cdots,n_{cp} \label{approx_MILP_thermal_effect_lower_cons}\\
& \displaystyle \sum_{i=1}^{n_{ch-lps}}\ell_i\overline{k_i^u}h_{is} \le \bar{f}+\mu_{s}^u & s=1,\cdots,n_{cp} \label{approx_MILP_thermal_effect_upper_cons}\\
&  \mu_s^l, \mu_s^u \ge 0  & s=1,\cdots,n_{cp} \label{approx_MILP_mu positivity cons}
\end{eqnarray}
\end{subequations}
\hrule
\end{figure*}

\section{A case study}
\label{sec:A case study}

\begin{figure*}[h!]
  \subfigure[]{%
    \includegraphics[width=0.5\textwidth]{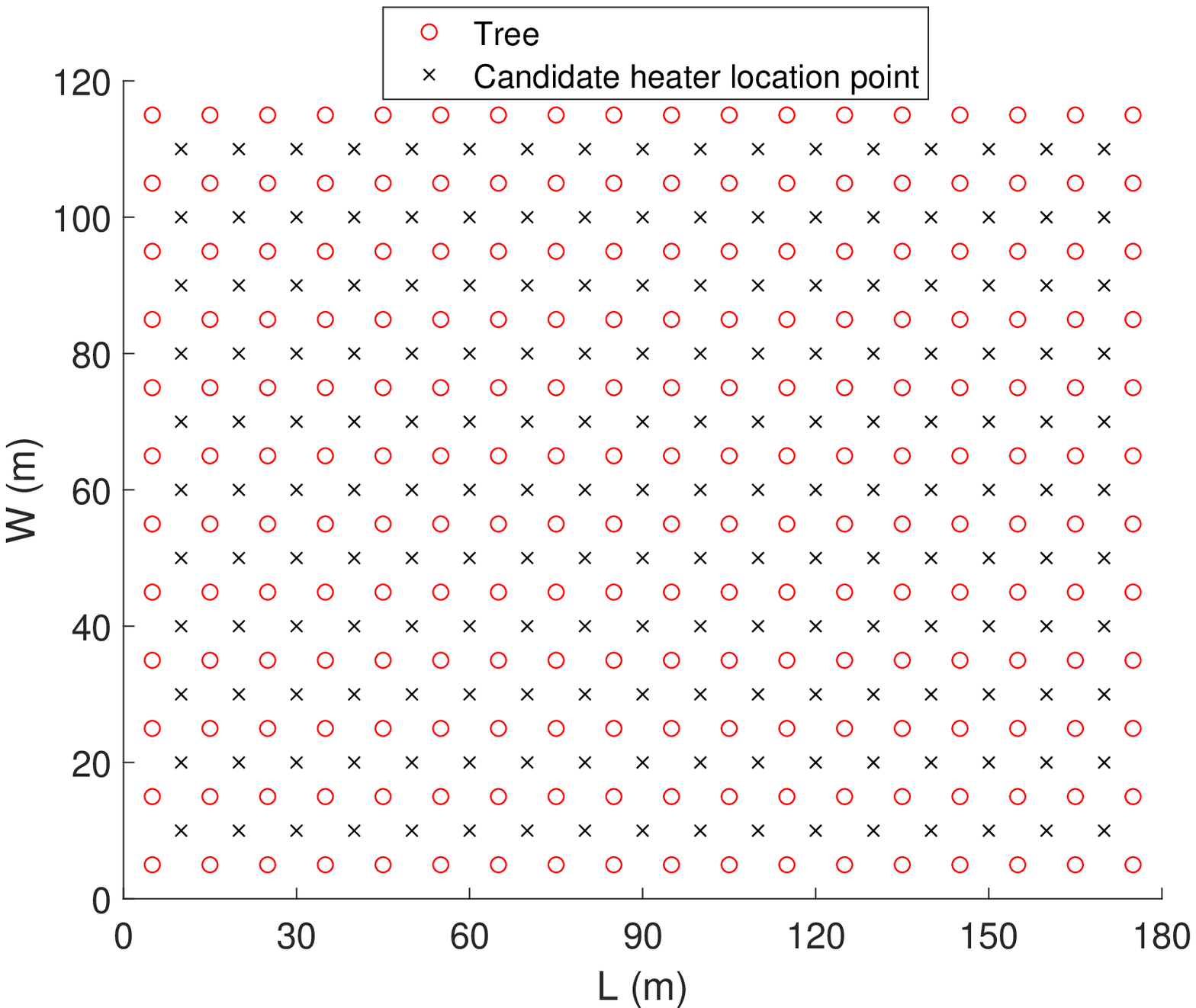} \label{fig:trees_cand_heaters}
  }
  \quad
  \subfigure[]{%
    \includegraphics[width=0.5\textwidth]{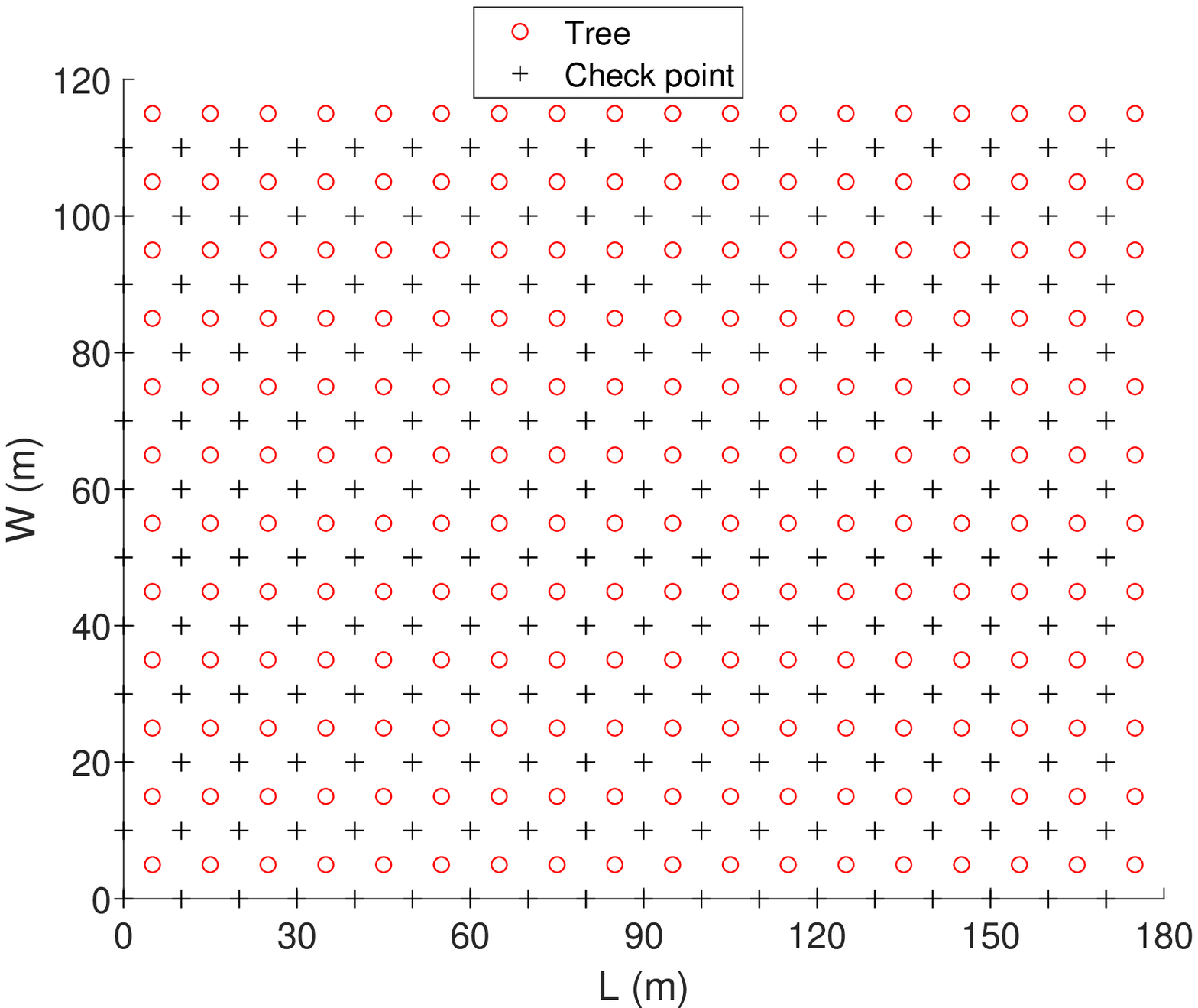} \label{fig:trees_check_point}
  }
  \caption{Distribution of trees, candidate heater locations and check points: (a) tree and candidate heater locations, (b) tree and check point locations.}
  \label{fig:case_study_ch_loc_cp}
\end{figure*}

\begin{figure*}[h!]
\centering
\includegraphics[scale=0.55]{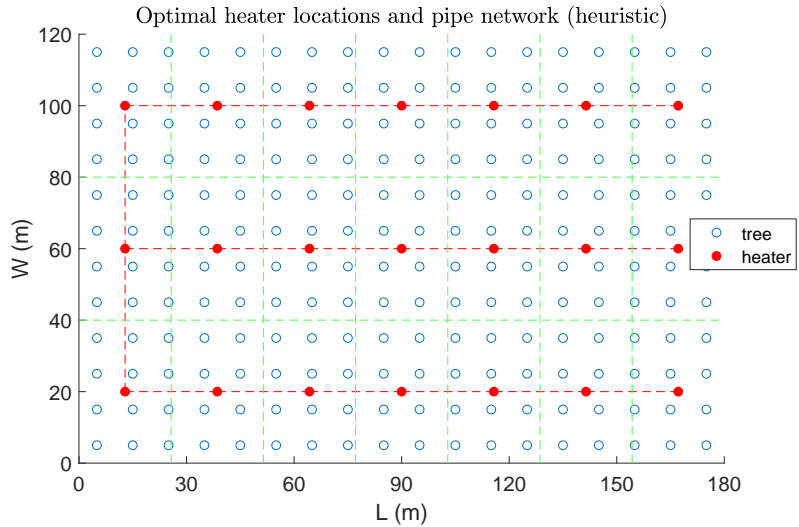}
\caption{Heuristic-based heater locations and pipe network.}
\label{fig:Intuition-based heater locations and pipe network}
\end{figure*}

In this section we will consider a case study and
compare the results from a heuristic-based  design
with the multi-objective robust optimization results.
The considered large-scale orchard together
with candidate heater location points and check points are given in Figure~\ref{fig:case_study_ch_loc_cp}
and its parameters are given in
Table~\ref{table:Orchard and optimization parameters}.
The heuristic-based  design
consists of two stages. In the
first stage, we divide the orchard into $k$
parts having
the same area and put a heater at the center of each part. If a heater is put too close to a tree (within $d_{ht}$ meters), it is pushed from the tree in the same direction so that the distance between the tree and the heater is $d_{ht}$ m. By doing so, we ensure that no heater is located too close to the trees.  In the second stage, we find pairwise
distances between the centers
to construct an undirected graph and then
a MST from the resulting graph using
Kruskal's algorithm.
The  heater locations and pipe network using this
heuristic are given in Figure \ref{fig:Intuition-based heater locations and pipe network}.

\begin{table}
\centering
\caption{Orchard and optimization parameters}
\label{table:Orchard and optimization parameters}
 \begin{tabular}{|p{1.5cm}||p{5cm}||l|}
   \hline
   \bf{parameter} &  \bf{description}  &\bf{value} \\ \hline \hline
    $L$ & orchard length (m)  & 180   \\  \hline
    $W$   & orchard width  (m)  & 120  \\  \hline
    $n_{t}$ & number of trees & 216 \\ \hline
   $k$ & number of heaters & 21 \\ \hline
    $n_{ch-lps}$ & number of candidate heater location points & 187  \\ \hline
    $n_{cp}$ & number of check points & 216 \\ \hline
   $d_{ht}$ & distance between the root of a tree and center of a heater (m) &  3   \\ \hline
   $\underbar{f}$ &  minimum total power fraction at check points & 0.5 \\ \hline
    $\bar{f}$ &  maximum  total power fraction at check points & 1  \\ \hline
   $k_i^u$  &   parameter to represent the uncertainty in the  heating effect of a heater
                  at a point & $[0.8, \, 1]$ \\ \hline
    $\underline{k_i^u}$  &  lower bound of $k_i^u$ &  0.8 \\ \hline
    $\overline{k_i^u}$   &  upper bound of $k_i^u$ &  1 \\ \hline
   $k_{tun}$ & tuning variable for spatial variation of the heating effect of heaters & 0.01 \\ \hline
    $\Delta_{t}$ &  horizontal and vertical distance between the roots of adjacent trees in the orchard (m)   &  10 \\ \hline
    $\Delta_{ch-lps}$ &  horizontal and vertical distance between  adjacent candidate heater locations in the orchard (m)  & 10   \\ \hline
   $\Delta_{cp}$ &  horizontal and vertical distance between adjacent check points in the orchard (m)  & 10 \\ \hline
   $\beta_1^{nor}$ & normalization parameter for the length of k-MST  & 600  \\ \hline
   $\beta_2^{nor}$ & normalization parameter for
   the sum of power range violations & 240  \\ \hline
 \end{tabular}
\end{table}

\begin{figure*}[h!]
\centering
\includegraphics[scale=0.5]{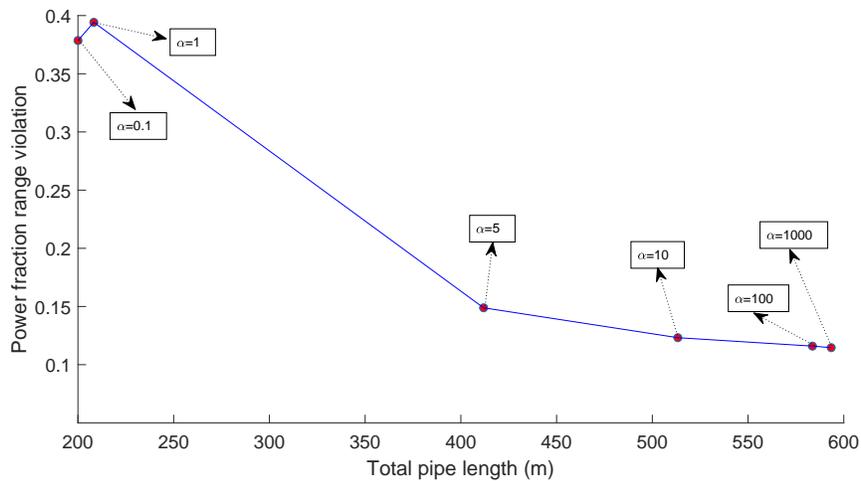}
\caption{Pareto plot for the  multi-objective robust  optimization problem.}
\label{fig:pareto_curve}
\end{figure*}

\begin{figure*}[ht!]
\centering
 \subfigure[]{%
    \includegraphics[width=0.475\textwidth]{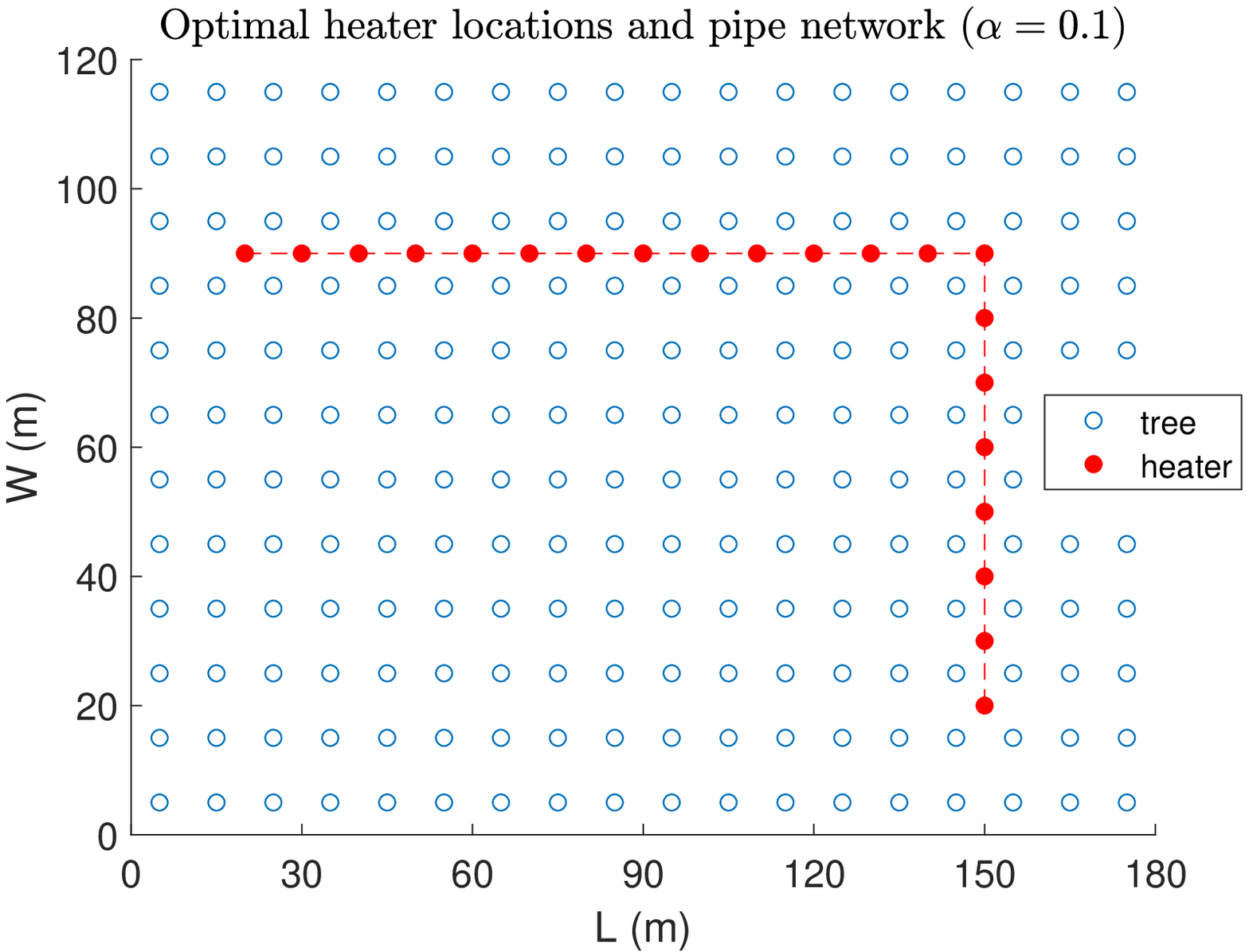}
  }
  \quad
 \subfigure[]{%
    \includegraphics[width=0.475\textwidth]{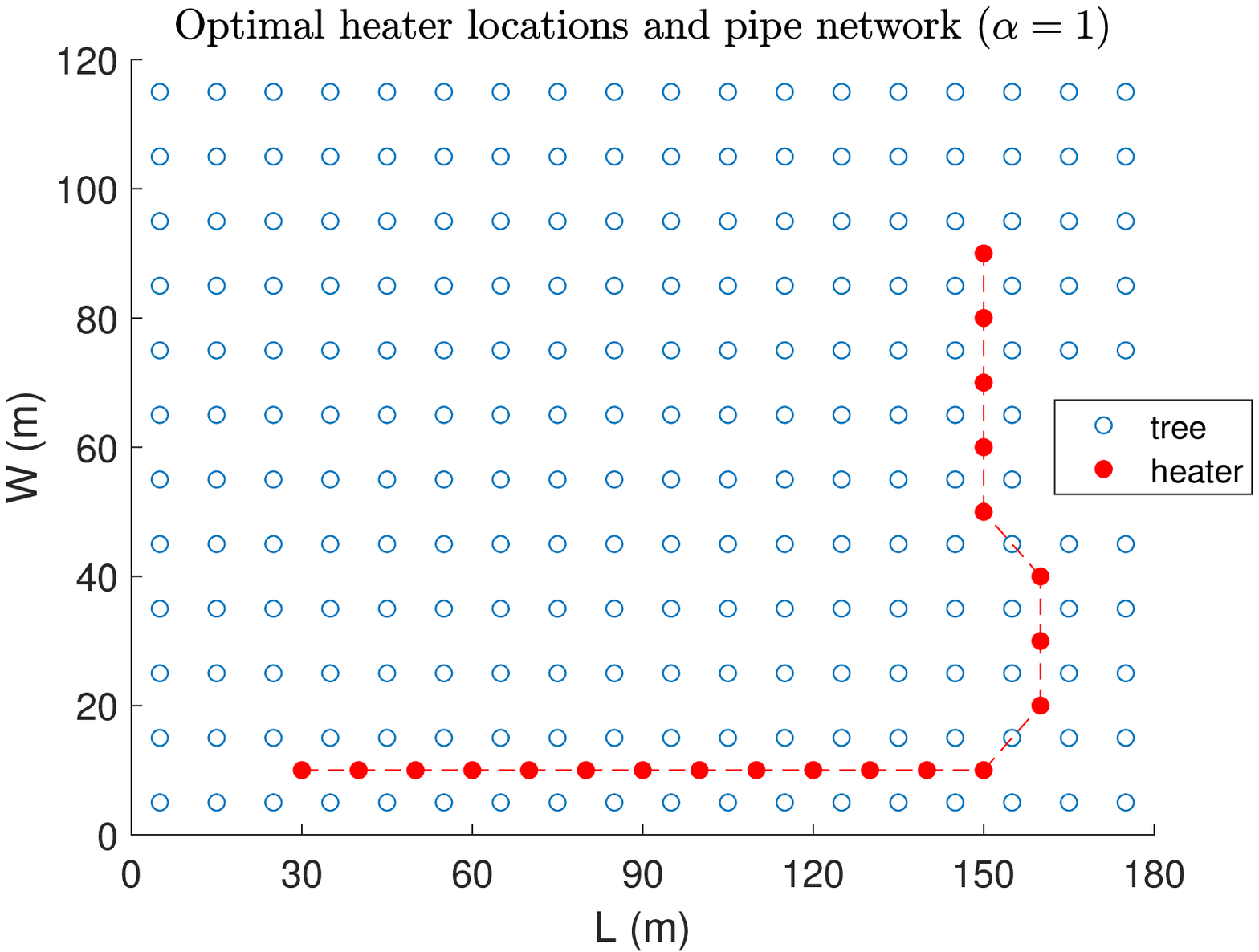}
  }
  \quad
   \subfigure[]{%
    \includegraphics[width=0.475\textwidth]{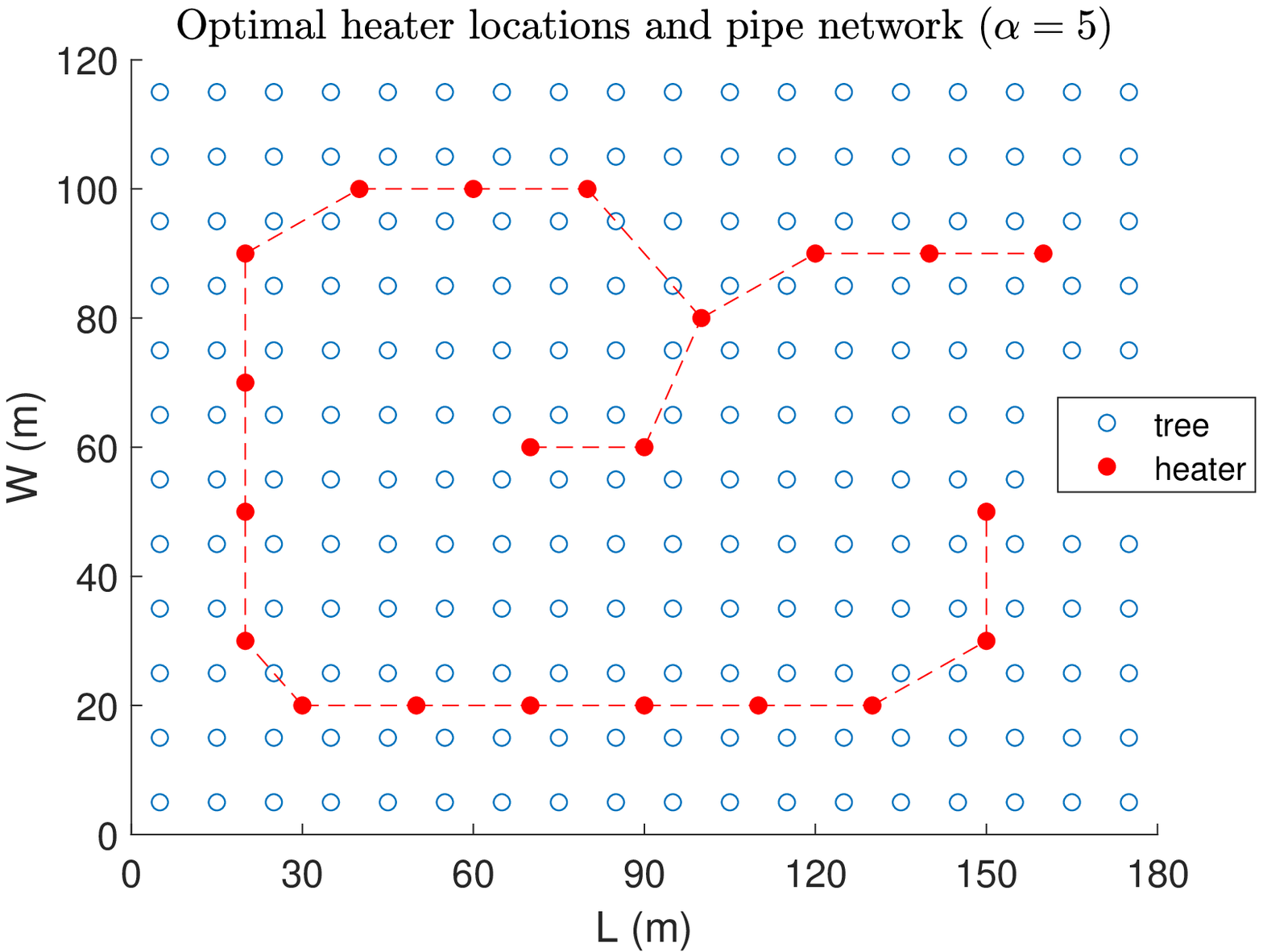}
  }
  \quad
  \subfigure[]{%
    \includegraphics[width=0.475\textwidth]{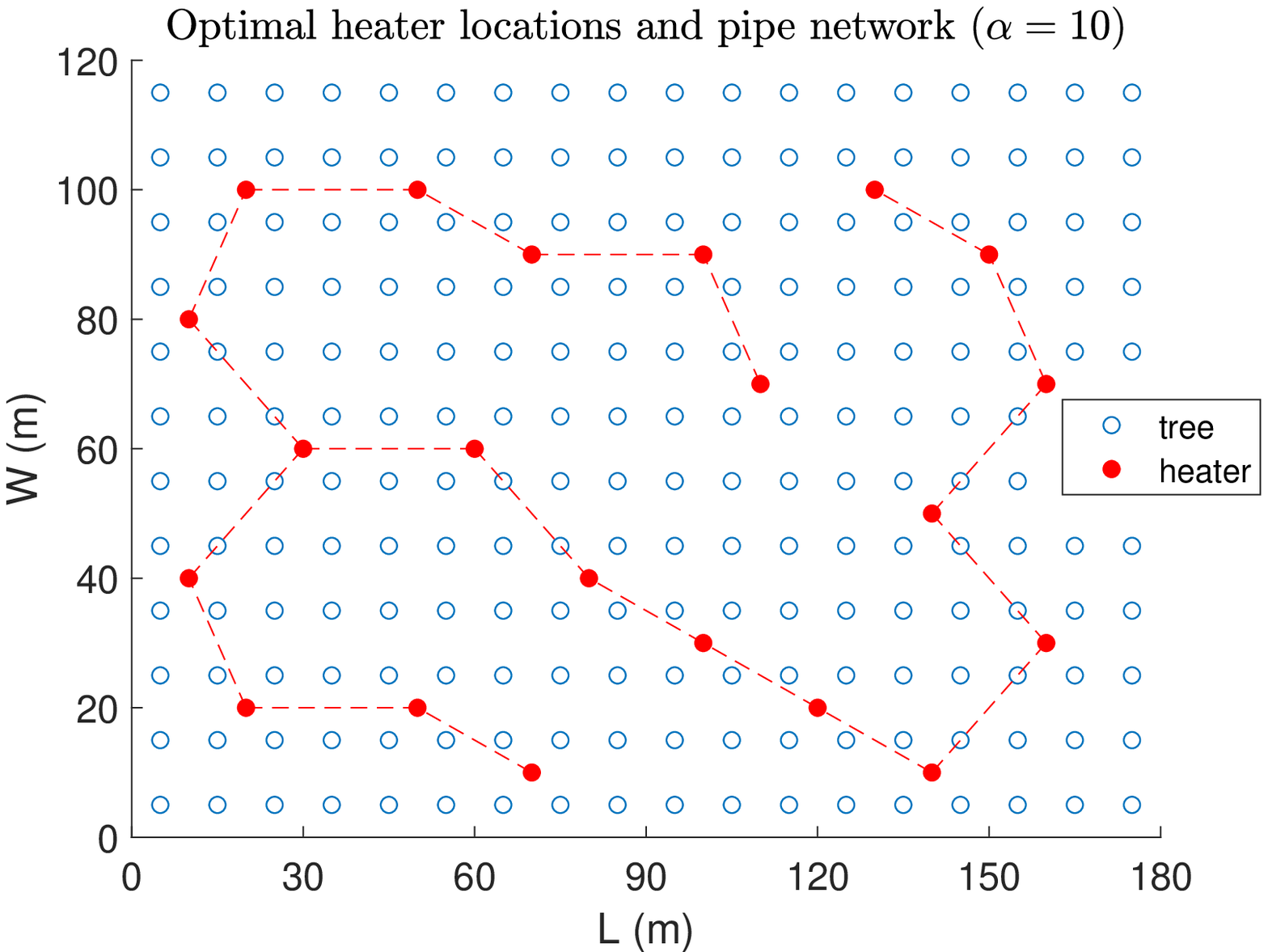}
  }
  \quad
 \subfigure[]{%
    \includegraphics[width=0.475\textwidth]{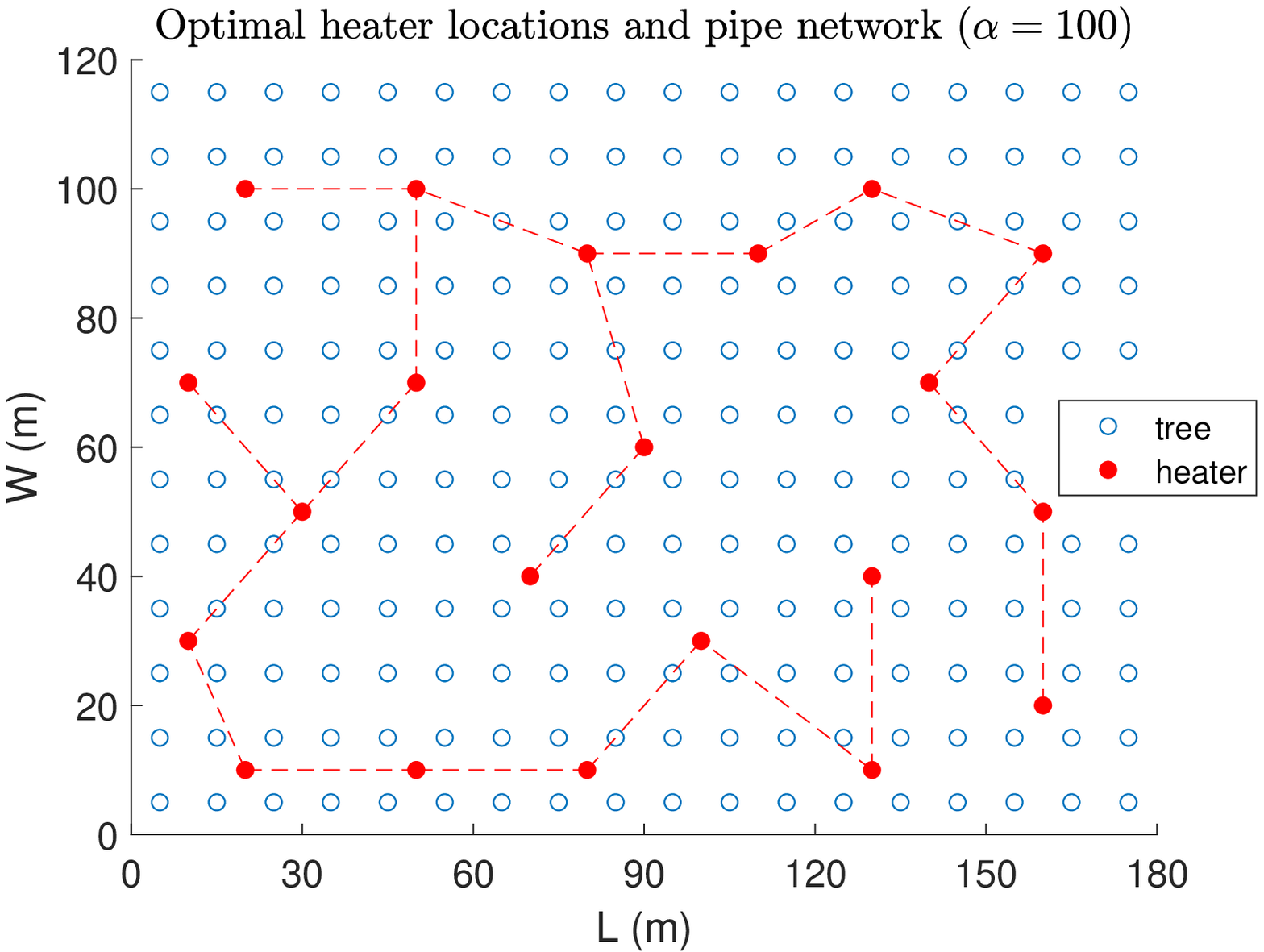}
  }
  \quad
  \subfigure[]{%
    \includegraphics[width=0.485\textwidth]{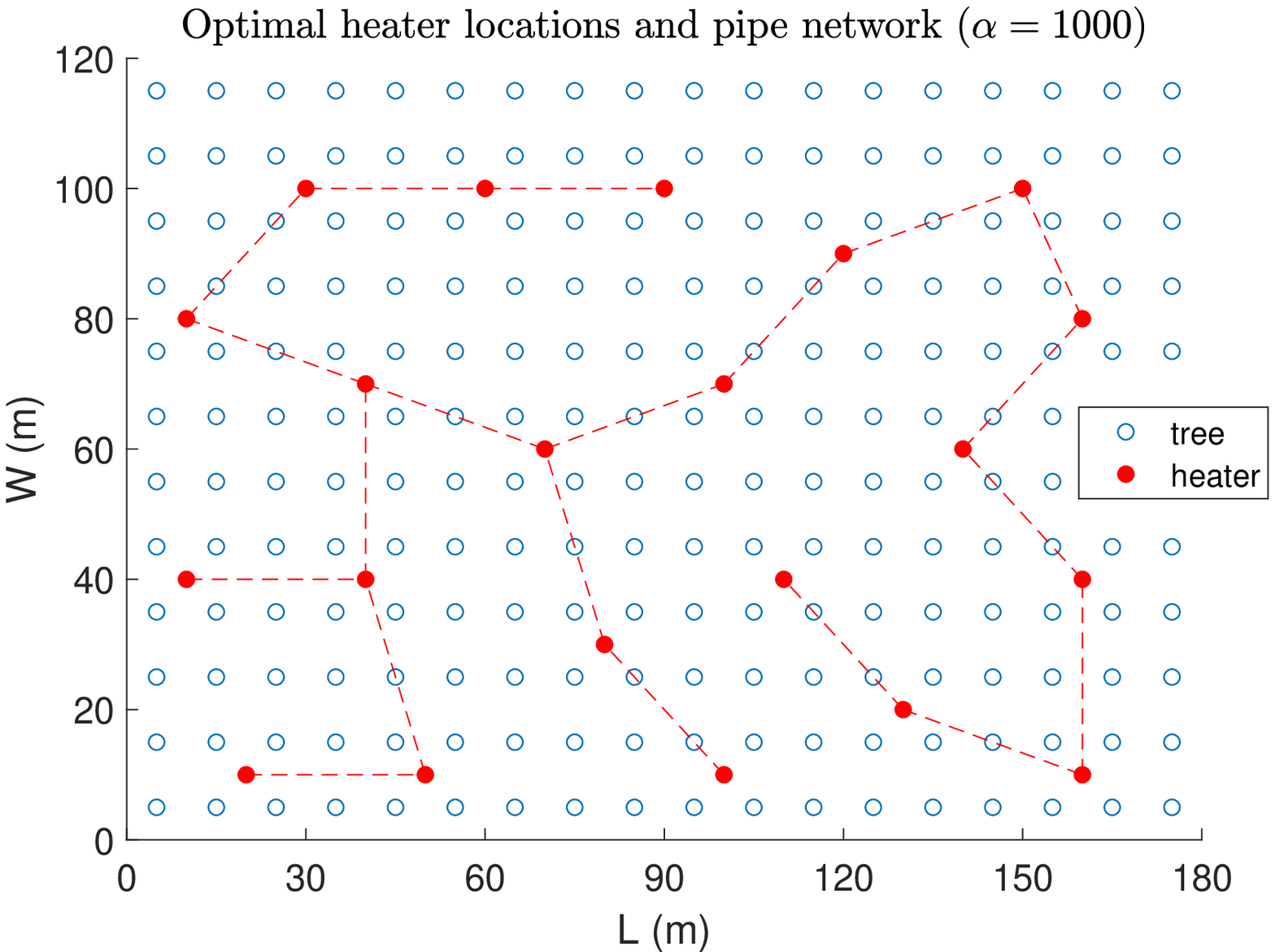}
  }
  \caption{Optimal heater locations and pipe network using multi-objective robust optimization-based design for a set of $\alpha$ values.}
  \label{fig:multi-objective optimization configs}
\end{figure*}

The multi-objective robust optimization results  were obtained using Gurobi\cite{gurobi_2021} as a solver on a laptop with the following hardware specifications: 8GB RAM, Intel(R) Core(TM) i7-8550U CPU  @ 1.80GHz 1.99 GHz. Solver time was set to 3 hours
to find a solution close to optimal.
The optimization results are given
in Table \ref{table:multi-objective optimization and intuition-based results} where the first column
includes a series of values for the multi-objective optimization parameter $\alpha$, the second column gives
the optimality gaps, and the last two columns
denote
\begin{align*}
& \text{obj}_\text{part1} \triangleq \sum_{e=\{i,j\} \in \mathcal{E} }z_{\{i,j\}}d_{i,j}^{ch-lp} \\
& \text{obj}_\text{part2} \triangleq \frac{1}{n_{cp}}\sum_{s=1}^{n_{cp}}(\mu_s^l+\mu_s^u)
\end{align*}
which are the total pipe length and average power
fraction range violation, respectively.
The Pareto curve of
the multi-objective  robust  optimization and
the corresponding optimal heater locations and pipe network
for each case of the considered $\alpha$ value
are given in Figure~\ref{fig:pareto_curve} and Figure~\ref{fig:multi-objective optimization configs}, respectively,
from which we conclude that among the considered
$\alpha$ values $\alpha=5$ seems a good choice with
$\text{obj}_\text{part1}=411.87$ (m) and $\text{obj}_\text{part2}=0.149$.
In the Pareto curve we notice that the point
corresponding to $\alpha=1$ is higher than the
point corresponding to $\alpha=0.1$. Normally,
the opposite should be the case. The cause for this
abnormality is the fact that the optimality gap corresponding to $\alpha=1$
($35.9 \%$)
is considerably higher than
the optimality gap corresponding to $\alpha=0.1$
($6.23 \%$)
obtained after 3 hours run of the optimization
algorithm (see, Table \ref{table:multi-objective optimization and intuition-based results}).

\begin{table}
\centering
\caption{Multi-objective optimization and heuristic-based results}
\label{table:multi-objective optimization and intuition-based results}
\begin{tabular}{|l|l|l|l|}
  \hline
  $\alpha$ & $ \text{optimality gap} (\%)$  & $\text{obj}_\text{part1}$ (m)  &  $\text{obj}_\text{part2} (-)$ \\ \hline\hline
  $0.1 $ & 6.23  & 200   & 0.379  \\ \hline
  $1 $ & 35.9  & 208,28   & 0.394  \\ \hline
  $\bf{5} $ & \bf{31.94}  & \bf{411.87}   &\bf{0.149}  \\ \hline
  $10$ & 24.56  & 513.31   & 0.123  \\ \hline
  $100$ & 25.40  & 583.64   & 0.116  \\ \hline
  $1000$ & 30.50  & 593.43   & 0.114  \\ \hline \hline
  \bf{Heuristic-based} & ----  & \bf{542.85}   & \bf{0.326}  \\ \hline
\end{tabular}
\end{table}

Table \ref{table:multi-objective optimization and intuition-based results} also shows the
heuristic-based design results. When calculating
the power fraction  range violations in the heuristic
case, $k_i^u$ value for each heater was created
as a random number in the range [0.8, 1]. When the
multi-objective robust optimization results corresponding
to the best case $\alpha=5$ are compared to
the heuristic-based results,
we have a 24.13\% reduction in the total pipe length
and a 54.29\% reduction in power fraction range violations, which clearly show  the savings provided
by the developed multi-objective
robust optimization approach.

\newpage
\section{Conclusions}
\label{sec:Conclusions}

Cost-effective frost prevention in large-scale orchards is a challenging problem, and development of
any cheap and widespread usable technology for it will have a huge societal impact. Being aware of the importance of this problem, in this paper, first we presented a renewable energy-integrated (solar thermal) solution for frost prevention in horticulture, thus providing an important potential application where renewable energy can be used for. Second, we developed a multi-objective robust MILP formulation
(i) to determine optimally the location of hot air blowers to heat the orchard
in a balanced way to achieve maximum frost protection, (ii) to optimize the layout of
hot air distribution pipe network to minimize the investment cost and energy losses.
The proposed optimization method involved, as part of it,
the development of a MILP-based k-MST formulation suitable for multi-objective optimization problems.

The proposed optimization method was tested on a case study with the dimensions $L \times W =180 \text{m} \times 120 \text{m}$.  The  optimization results
corresponding to the the optimal value of
$\alpha$ obtained from  the Pareto plot
and a 3 hours run of the algorithm
were compared with those of a heuristic-based design to show the degree of  savings, which are a 24.13\% reduction in the total pipe length
and a 54.29\% reduction in power fraction range violations.
These savings clearly illustrate the importance of the use of
a numerical optimization framework in optimal design of the suggested hybrid energy frost prevention system. Although here we focused on frost prevention
for horticulture, the proposed framework can be used with minor modifications for other agricultural applications as well.

Examples of future works are
economic feasibility analysis of the proposed hybrid energy solution,
determination of the minimal number of heaters to be used for a given
large-scale orchard and integration of a decomposition
scheme for  very large-scale optimization problems corresponding to very large-scale orchards.

\bibliographystyle{IEEEtran}

\begin{IEEEbiography}[{\includegraphics[width=0.9in,height=1.2in,clip]{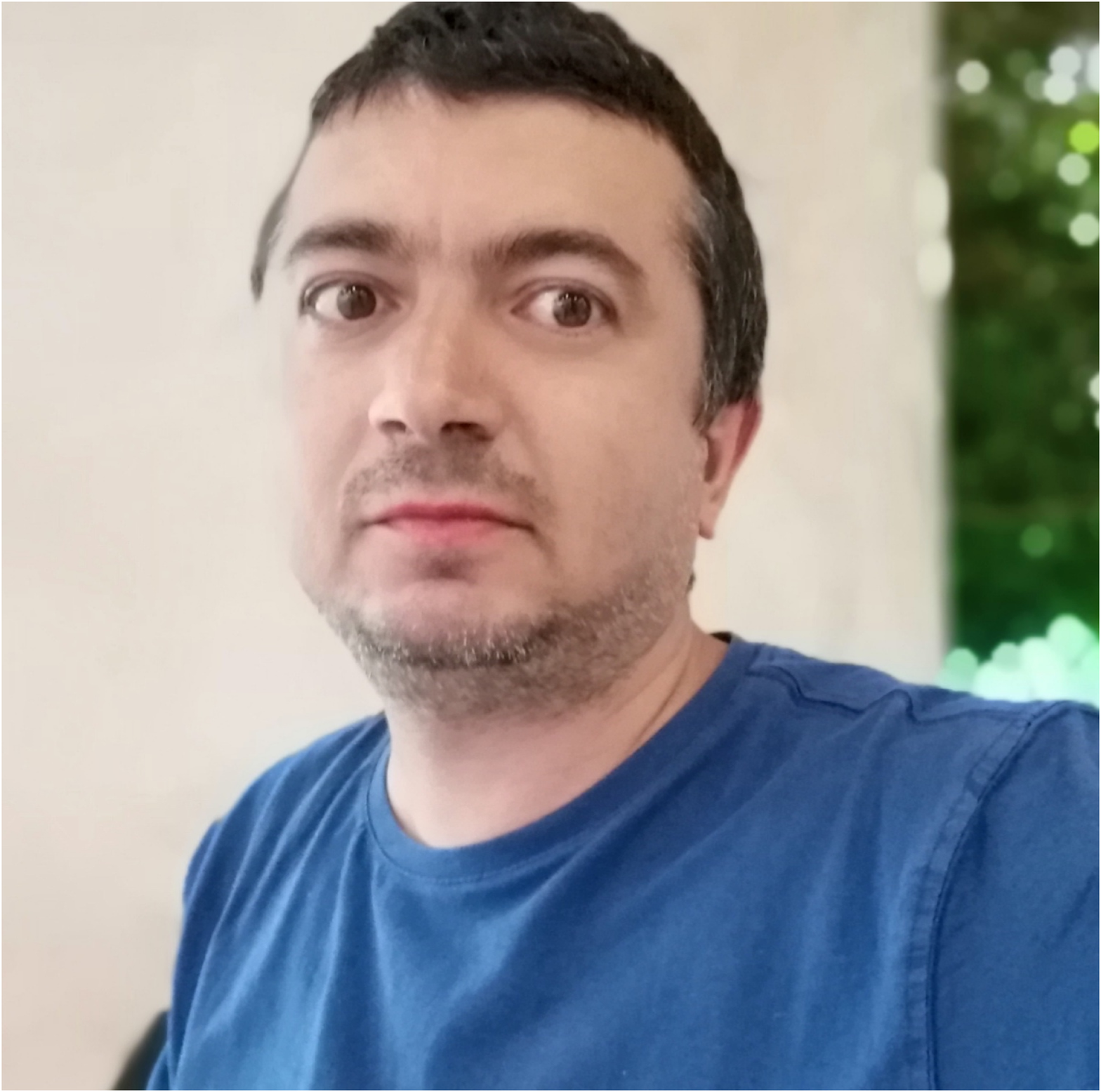}}]{\textbf{Ercan Atam}}
received his PhD  from Bo\u{g}azi\c{c}i University,
Istanbul, Turkey, in 2010.
He was a Postdoctoral
Researcher with LIMSI-CNRS, France,
from 2010 to 2012 working on fluid flow control.
From 2012 to 2015, he was a Postdoctoral
Researcher with KU Leuven, Belgium,
working on control and optimization
for smart buildings.
Currently, he is an honorary
research fellow at Imperial College London, UK. His research interests
include optimization, predictive control, machine learning,
renewable energy-integrated co-design applications and
smart horticulture.
\end{IEEEbiography}

\begin{IEEEbiography}[{\includegraphics[width=0.9in,height=1.2in,clip]{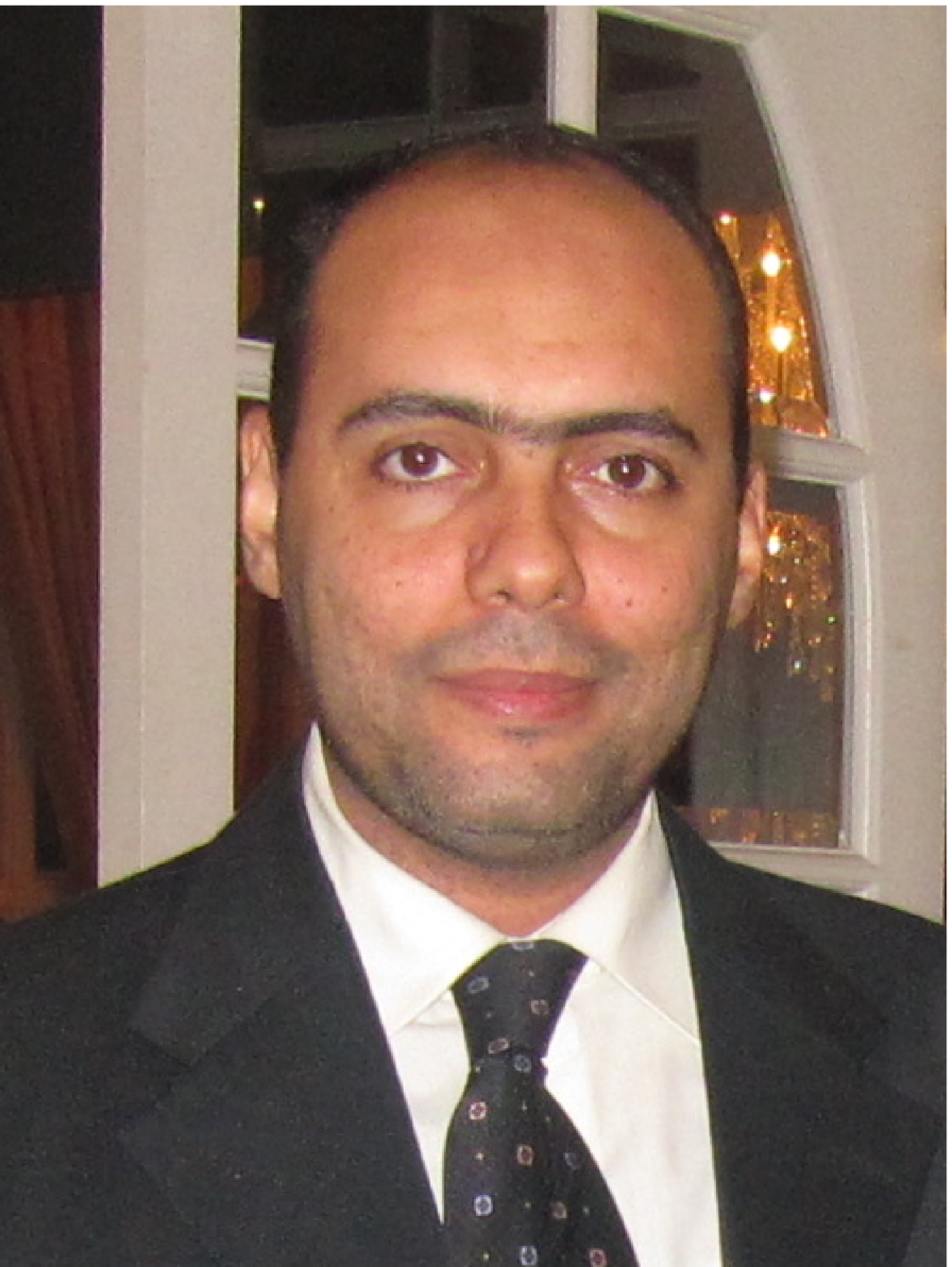}}]{\textbf{Tamer F. Abdelmaguid}}
received his Ph.D. in Industrial and Systems Engineering at the University of Southern California in 2004. Currently, he is a professor of industrial engineering at Cairo University, Egypt. He has served as a visiting assistant professor at King Saud University from 2008 to 2010, and as a visiting associate professor at the American University in Cairo from 2017 to 2019. His research interests are related to mathematical modeling and optimization of manufacturing processes and production planning and scheduling.\end{IEEEbiography}

\begin{IEEEbiography}[{\includegraphics[width=0.9in,height=1.2in,clip]{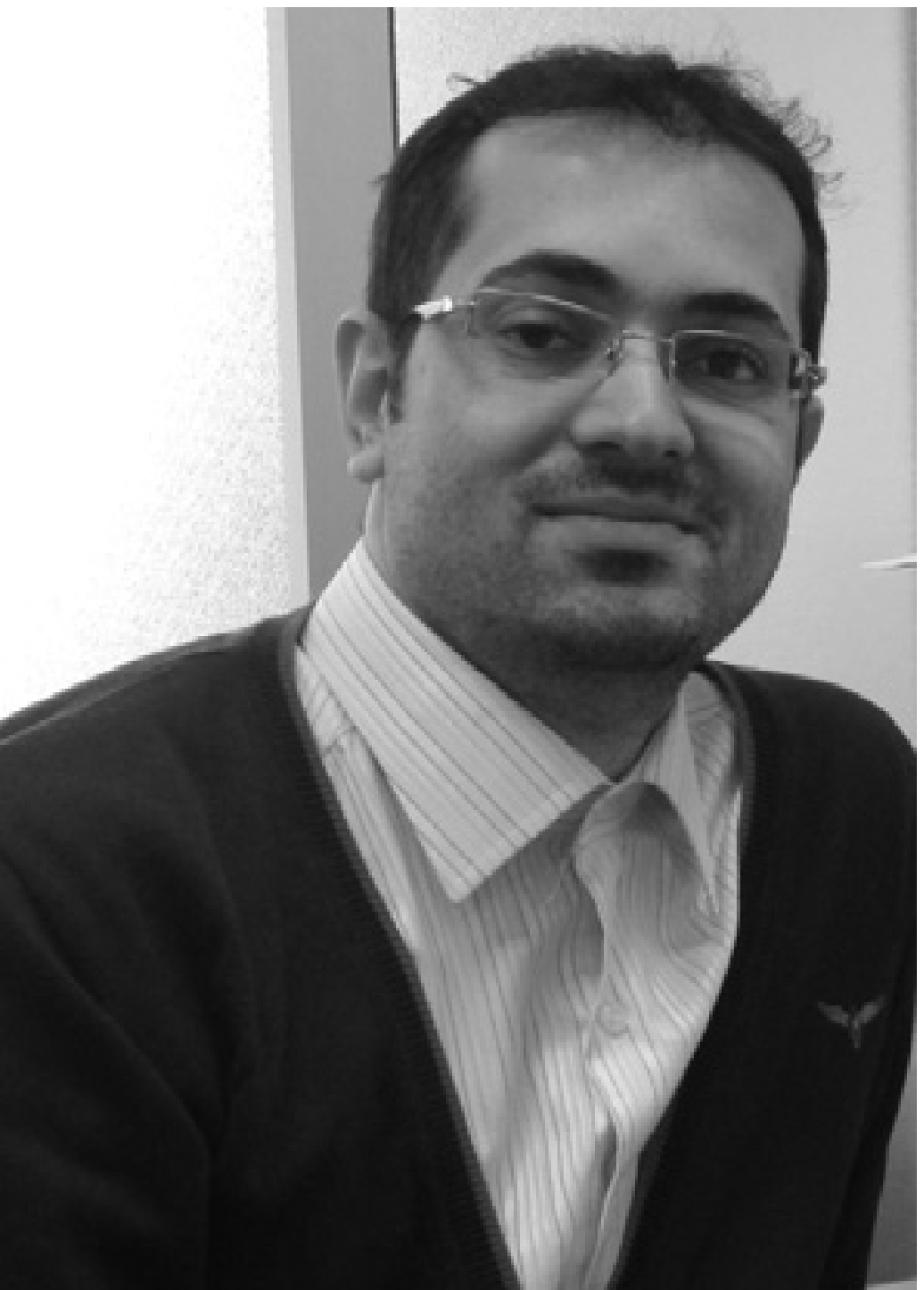}}]{\textbf{Muhammed Emre Keskin}}
received his B.S. degree from Industrial Engineering department of Bilkent University, Turkey, in 2004 and his M.S. and Ph.D. degrees from Industrial Engineering department of Bo\u{g}azi\c{c}i University respectively in 2007 and 2014. He is currently working as an associate professor at Atat\"{u}rk University. His research interests are mobile wireless sensor networks, mixed integer linear programming and decomposition of integer programs.
\end{IEEEbiography}

\begin{IEEEbiography}[{\includegraphics[width=1in,height=1.2in,clip]{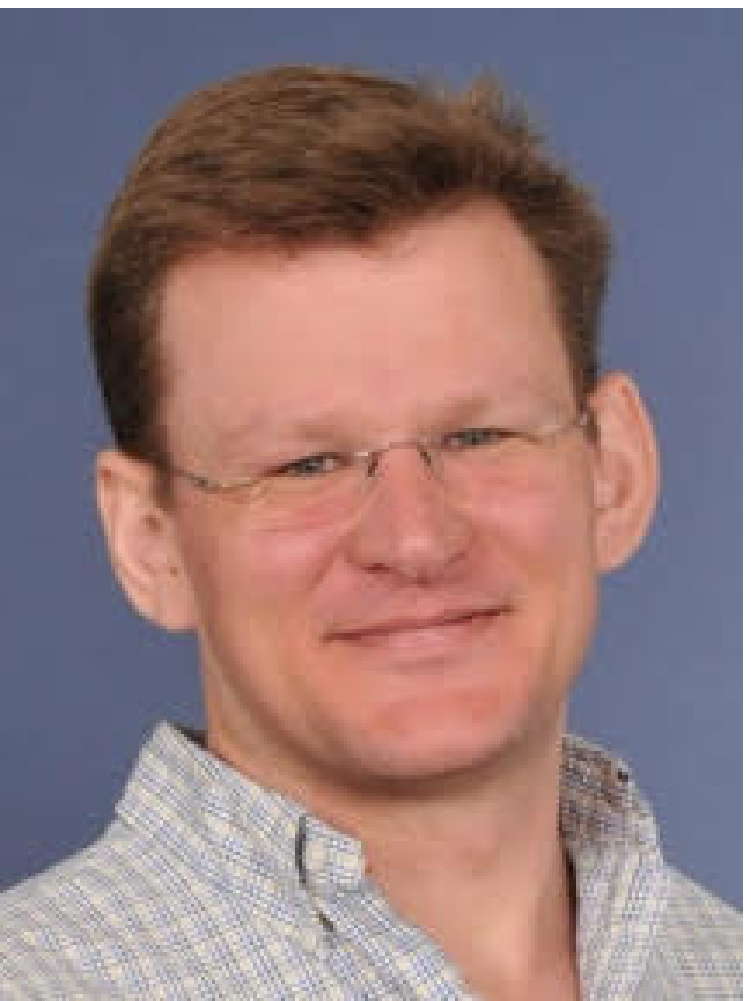}}]{\textbf{Eric C. Kerrigan}}
(S'94-M'02-SM'16) received a BSc(Eng) from the University of Cape Town and a PhD from the University of Cambridge. His research is in the design of efficient numerical methods and computing architectures for solving  optimal control problems in real-time, with applications in the design of aerospace, renewable energy and information systems. He is  the chair of the IFAC Technical Committee on Optimal Control, a Senior Editor of IEEE Transactions on Control Systems Technology and  an Associate Editor of IEEE Transactions on Automatic Control and the European Journal of Control.

\end{IEEEbiography}

\end{document}